\documentclass[aps,pra,twocolumn,showpacs,superscriptaddress,floatfix]{revtex4-1}

\usepackage{amsmath,amssymb}
\usepackage{graphicx}
\usepackage{color}
\usepackage{rotating}
\usepackage{bm}
\usepackage{setspace}
\usepackage{hyperref}
\usepackage{multirow}

\hypersetup{
colorlinks=true,
urlcolor= blue,
citecolor=blue,
linkcolor= blue,
bookmarks=true,
bookmarksopen=false,
}

\begin{document}

\title{Unusual Valence State in the Antiperovskites Sr$_3$SnO and Sr$_3$PbO Revealed by X-ray Photoelectron Spectroscopy}

\author{D. Huang}
\email{d.huang@fkf.mpg.de}
\affiliation{Max Planck Institute for Solid State Research, 70569 Stuttgart, Germany}
\author{H. Nakamura}
\email{hnakamur@uark.edu}
\altaffiliation{Present address: Department of Physics, University of Arkansas, Fayetteville, Arkansas 72701, USA}
\affiliation{Max Planck Institute for Solid State Research, 70569 Stuttgart, Germany}
\author{K. K\"{u}ster}
\affiliation{Max Planck Institute for Solid State Research, 70569 Stuttgart, Germany}
\author{A. Yaresko}
\affiliation{Max Planck Institute for Solid State Research, 70569 Stuttgart, Germany}
\author{D. Samal}
\affiliation{Institute of Physics, Bhubaneswar 751005, India}
\affiliation{Homi Bhabha National Institute, Mumbai 400085, India}
\author{N. B. M. Schr\"{o}ter}
\affiliation{Swiss Light Source, Paul Scherrer Institute, CH-5232 Villigen PSI, Switzerland}
\author{V. N. Strocov}
\affiliation{Swiss Light Source, Paul Scherrer Institute, CH-5232 Villigen PSI, Switzerland}
\author{U. Starke}
\affiliation{Max Planck Institute for Solid State Research, 70569 Stuttgart, Germany}
\author{H. Takagi}
\affiliation{Max Planck Institute for Solid State Research, 70569 Stuttgart, Germany}
\affiliation{Institute for Functional Matter and Quantum Technologies, University of Stuttgart, 70569 Stuttgart, Germany}
\affiliation{Department of Physics, University of Tokyo, 113-0033 Tokyo, Japan}

\date{\today}

\begin{abstract}
The class of antiperovskite compounds $A_3B$O ($A$ = Ca, Sr, Ba; $B$ = Sn, Pb) has attracted interest as a candidate 3D Dirac system with topological surface states protected by crystal symmetry. A key factor underlying the rich electronic structure of $A_3B$O is the unusual valence state of $B$, i.e., a formal oxidation state of $-4$. Practically, it is not obvious whether anionic $B$ can be stabilized in thin films, due to its unusual chemistry, as well as the polar surface of $A_3B$O, which may render the growth-front surface unstable. We report X-ray photoelectron spectroscopy (XPS) measurements of single-crystalline films of Sr$_3$SnO and Sr$_3$PbO grown by molecular beam epitaxy (MBE). We observe shifts in the core-level binding energies that originate from anionic Sn and Pb, consistent with density functional theory (DFT) calculations. Near the surface, we observe additional signatures of neutral or cationic Sn and Pb, which may point to an electronic or atomic reconstruction 
with possible impact on putative topological surface states.
\end{abstract}


\maketitle


\section{Introduction}

Complex oxides have long provided a rich platform to explore exotic electronic phases that emerge from the interplay of charge, spin and orbital degrees of freedom~\cite{Imada_RMP_1998}. In recent years, efforts to engineer Dirac, Weyl and other topological semimetallic phases in these compounds have intensified~\cite{Uchida_JPD_2018}. The effects of strong electronic correlations~\cite{Fujioka_NatCommun_2019}, magnetism~\cite{Wan_PRB_2011} and interface reconstructions~\cite{Hwang_NatMat_2012} in complex oxides are expected to enrich the topological phases that can be realized. Such investigations are facilitated by the ability to synthesize these compounds in thin-film heterostructures.

\begin{figure}
\includegraphics[scale=1]{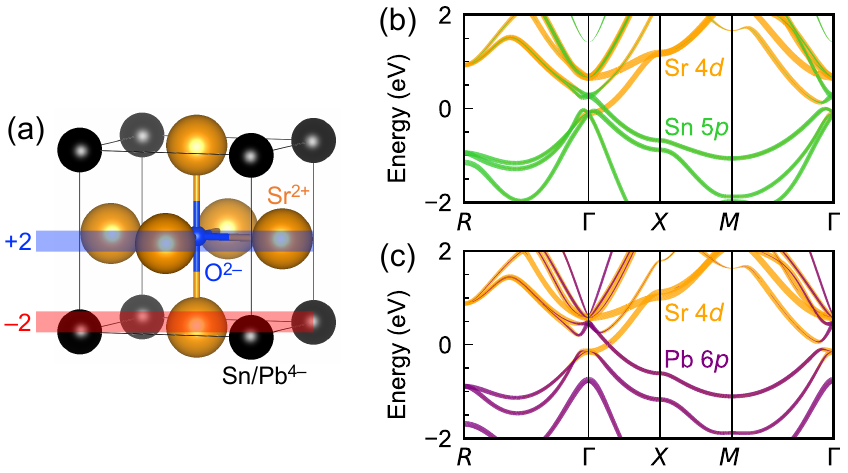}
\caption{(a) Crystal structure of the antiperovskite Sr$_3$(Sn, Pb)O. The horizontal bars (red and blue) illustrate polar (001) planes. (b), (c) Band structure plots of Sr$_3$SnO and Sr$_3$PbO. The thickness of the orange (green/purple) line denotes the weight of the projection of the given state onto the Sr 4$d$ (Sn 5$p$/Pb 6$p$) orbitals.}
\label{Fig1}
\end{figure}

A pertinent example is the class of antiperovskites (or inverse perovskites) with chemical formula $A_3B$O, where $A$ is an alkaline earth metal (Ca, Sr or Ba) and $B$ is Sn or Pb. These compounds crystallize into the archetypal perovskite structure, but with the usual positions of the cations and anions exchanged [Fig.~\ref{Fig1}(a)]. These antiperovskites have been predicted to host a unique set of electronic properties. According to rigorous classification, several members of this family are topological crystalline insulators~\cite{Hsieh_PRB_2014} with type-I and type-II Dirac surface states~\cite{Chiu_PRB_2017}. However, the actual band gap, which lies along the $\Gamma$-$X$ line at six equivalent points in the Brillouin zone (BZ), is only a few tens of millielectronvolts, such that in the vicinity of these points, there is a quasilinear 3D Dirac dispersion~\cite{Kariyado_JPSJ_2011, Kariyado_JPSJ_2012, Kariyado_PRM_2017}. Experimentally, angle-resolved photoemission spectroscopy~\cite{Obata_PRB_2017}, magnetotransport~\cite{Suetsugu_PRB_2018, Obata_PRB_2019} and nuclear magnetic resonance~\cite{Kitagawa_PRB_2018} measurements have probed the possible 3D Dirac nature of the electrons in these compounds. Experiments have also revealed signatures of ferromagnetism arising from oxygen vacancies~\cite{Lee_APL_2013, Lee_MRS_2014}, high thermoelectric performance~\cite{Okamoto_JAP_2016}, superconductivity arising from Sr vacancies~\cite{Oudah_NatComm_2016, Oudah_SciRep_2019} and weak antilocalization due to spin-orbital entanglement~\cite{Nakamura_arXiv_2018}.

The rich electronic properties of the antiperovskites take as their fundamental origin the unusual valence state of $B$ (= Sn, Pb). In the ionic limit, the constituent elements of $A_3B$O would exist in the following oxidation states: $A^{2+}$, $B^{4-}$ and O$^{2-}$. We note that Bader analysis reveals that the effective charge of $B$ lies closer to $-$2 (see Section~\ref{secDFT}). Nevertheless, such a highly anionic state of $B$ implies that a large fraction of its outermost $p$-orbitals are occupied. This configuration produces an unusual situation in $A_3B$O, wherein the valence bands near the Fermi energy are dominated by $B$ $p$-orbitals and the conduction bands near the Fermi energy are dominated by $A$ $d$-orbitals (Figs.~\ref{Fig1}(b), (c); refer to Section~\ref{SecMet} for details of the band structure calculations). Around $\Gamma$, there is a moderate inversion between the $B$ $p$-bands and $A$ $d$-bands. When interorbital hybridization and spin-orbit coupling are taken into account, the six equivalent band crossings at the Fermi energy are only slightly gapped, resulting in the approximate 3D Dirac semimetallic phase~\cite{Kariyado_JPSJ_2011, Kariyado_JPSJ_2012, Kariyado_PRM_2017}, as well as the topological crystalline insulating phase in some cases~\cite{Hsieh_PRB_2014}.

It is natural to ask whether anionic Sn or Pb can be actually stabilized in a thin film. Not only do these anionic states represent unusual chemistry, resulting in extreme air sensitivity, but the antiperovskites may also be prone to surface reconstruction. As illustrated in Fig.~\ref{Fig1}(a), the (001) planes alternate between an overall oxidation state of $+2$ and $-2$, leading to a polar catastrophe at a surface~\cite{Ohtomo_Nature_2004}. To alleviate a divergence in the electrostatic potential, it is possible that Sr vacancies form at the surface (analogous to O vacancies in oxide perovskites), and/or Sn and Pb shift to a more stable valence state (neutral or cationic) via electronic reconstruction. If so, this has profound implications on surface states~\cite{Chiu_PRB_2017}, similar to the case of the Kondo insulator SmB$_6$, whose polar surface has complicated the elucidation of its topological properties~\cite{Zhu_PRL_2013}.

Previous measurements of bulk Sr$_{3-x}$SnO crystals uncovered signatures of anionic Sn~\cite{Oudah_SciRep_2019}. Using $^{119}$Sn M\"{o}ssbauer spectroscopy, Oudah \textit{et al.} observed an isomer shift of the main peak by $+$1.88 mm/s, matching that of Mg$_2$Sn, another compound in which Sn is formally $-$4. The situation in thin films, however, is less clear. Minohara \textit{et al.} performed X-ray photoelectron spectroscopy (XPS) measurements of Ca$_3$SnO films under ultra-high vacuum (UHV)~\cite{Minohara_JCG_2018}. The reported Sn 3$d_{5/2}$ spectrum showed a surface component corresponding to Sn$^{4+}$ or Sn$^{2+}$, as well as a bulk component, which they attributed to the antiperovskite phase. However, the bulk component had a binding energy of 484.8 eV, lying within the range expected for neutral Sn: 484.3-485.2 eV~\cite{NIST}. Further investigation is needed to clarify the anionic state of Sn in thin films.

Here, we performed XPS measurements of Sr$_3$SnO and Sr$_3$PbO films grown by molecular beam epitaxy (MBE) and kept in UHV conditions. In the bulk, we observe peaks in the Sn 3$d$ and Pb 4$f$ core levels that lie at lower binding energies than those of cationic or neutral Sn and Pb. DFT calculations confirm that these shifts match predictions for anionic Sn (Pb) in Sr$_3$SnO (Sr$_3$PbO). At the surface, we find signatures of cationic and neutral Sn and Pb, consistent with the scenario of an atomic or electronic reconstruction at the surface.

\section{Methods}
\label{SecMet}

Films of Sr$_3$SnO and Sr$_3$PbO with thickness $\sim$100 nm were grown in an Eiko MBE chamber with base pressure in the low $10^{-9}$ mbar range. The films were deposited on (001)-cut substrates of yttria-stabilized zirconia (YSZ), which were pre-coated at two opposite edges with Au or Nb for electrical grounding in XPS measurements. Elemental sources of Sr (99.9\% purity from vendor, further refined in house by sublimation), Sn (99.999\% purity) and Pb (99.999\%) were thermally sublimated from effusion cells. A mixture of 2\% O$_2$ in Ar gas was supplied through a leak valve (pressure range: $10^{-6}$ to $10^{-5}$ mbar). Since the samples reported in this work were grown at different times spanning a two-year period, different growth parameters were used in the course of optimizing film quality (example parameters can be found in Refs.~\cite{Samal_APLM_2016, Nakamura_arXiv_2018}). We will not focus on these systematic differences, but instead on the ubiquitous observation of an antiperovskite phase via XPS. 

Following growth, the films were examined \textit{in situ} using reflection high-energy electron diffraction (RHEED), then transferred in vacuum suitcases (Ferrovac GmbH; pressure range: low $10^{-10}$ mbar) for XPS measurements. We note that other films grown with identical conditions to these ones were capped with Au or Apiezon-N grease in an Ar glove box then characterized by X-ray diffraction (XRD) and/or transport~\cite{Samal_APLM_2016, Nakamura_arXiv_2018}. 

XPS data were acquired at the Max Planck Institute for Solid State Research (MPI-FKF) in a system equipped with a commercial Kratos AXIS Ultra spectrometer and a monochromatized Al K$_{\alpha}$ source (photon energy: 1486.6 eV). The base pressure was in the low $10^{-10}$ mbar range. An analyzer pass energy of 20 eV was used to collected detailed spectra. In addition to the antiperovskite films, we measured reference spectra from a Sn film (grown using MBE, transported in a vacuum suitcase) and a Pb foil (cleaned \textit{in situ} using Ar sputtering). The Sn film [Fig.~\ref{Fig3}(b)] showed charging due to poor electrical grounding; in this instance, we used the Fermi edge to recalibrate the binding energy. We also performed low-energy electron diffraction (LEED) on our films in an adjoining chamber equipped with a commercial SPECS ErLEED 150 system.

XPS spectra were analyzed using the CasaXPS software. To fit the various peaks, we used multiple Gaussian-Lorentzian mixture functions on top of a Shirley background. To constrain our fitting parameters, we fixed the doublet spacing energy of Sn 3$d$, Sr 3$d$ and Pb 4$f$ to their literature values of 8.41 eV, 1.79 eV and 4.86 eV, respectively~\cite{Moulder_1992}. We also constrained the area ratio of the doublets to 2:3 for $d$-core levels and 3:4 for $f$-core levels.

We also collected XPS data at grazing emission, which are more sensitive to the surface elemental composition. This allowed us to disentangle surface and bulk contributions in the XPS spectra. Similarly, for films measured at the ADRESS beamline of the Swiss Light Source (SLS)~\cite{Strocov_JSR_2010, Strocov_JSR_2014}, we were able to control the surface sensitivity by tuning the photon energy~\cite{SM}.

We performed DFT calculations using the Vienna \textit{ab-initio} simulation package (\texttt{VASP})~\cite{Kresse_CMS_1996, Kresse_PRB_1996}, which implements the projector augmented-wave (PAW) method~\cite{Bloch_PRB_1994, Kresse_PRB_1999}. The following electrons were treated as valence: $3s3p4s$ in Ca, $4s4p5s$ in Sr, $5s5p6s$ in Ba, $5s4d5p$ in Sn, $6s5d6p$ in Pb and $2s2p$ in O. We used the generalized gradient approximation (GGA) as parameterized by Perdew, Burke and Ernzerhof (PBE)~\cite{Perdew_PRL_1996}. An energy cutoff of 750 eV was used, along with a BZ sampling as dense as $28 \times 28 \times 28$ for the self-consistent calculation of the charge density. For the band structure calculations shown in Figs.~\ref{Fig1}(b) and (c), spin-orbit coupling was included in an additional non-self-consistent cycle. We also performed Bader charge analysis. Estimates of core-level shifts, which we performed using both \texttt{VASP} and \texttt{PY LMTO}, will be discussed in Section~\ref{secDFT}. Atomic structures were visualized using \texttt{VESTA}~\cite{Momma_JAC_2011}.

\section{Results and Discussion}

\subsection{RHEED and LEED}

\begin{figure}
\includegraphics[scale=1]{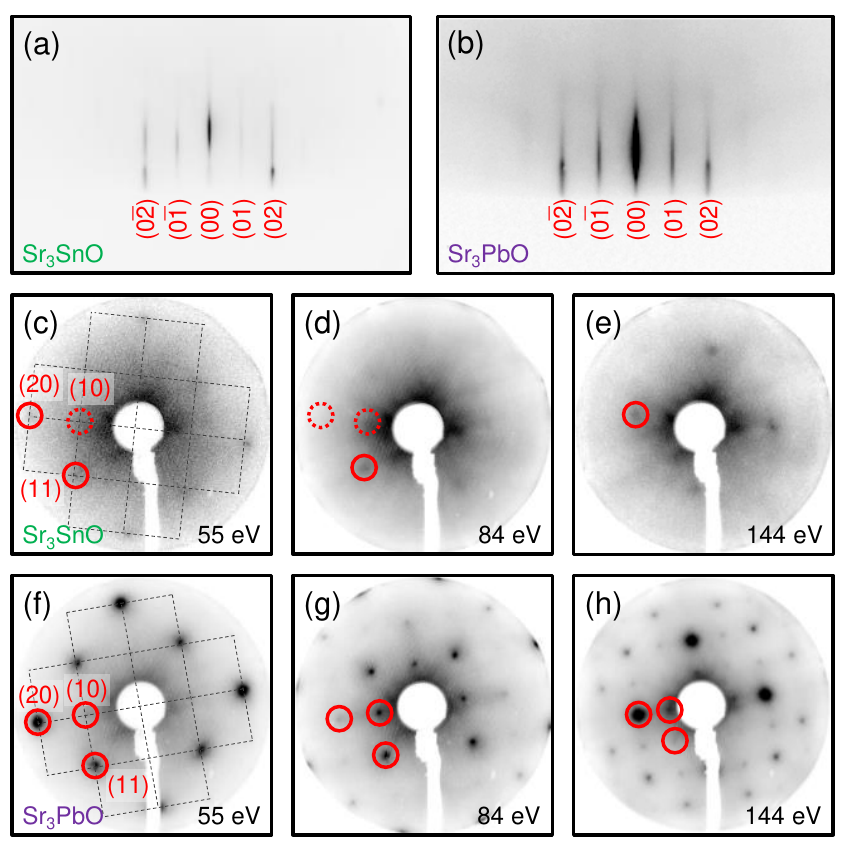}
\caption{(a), (b) RHEED images of Sr$_3$SnO (sample SS91) and Sr$_3$PbO (sample AP389) taken along the [100] direction. Electron energy: 15 keV. (c)-(e) LEED images of Sr$_3$SnO (sample SS60), acquired at 55 eV, 84 eV and 144 eV. (f)-(h) LEED images of Sr$_3$PbO (sample AP149), acquired at the same energies.}
\label{Fig2}
\end{figure}

Figures~\ref{Fig2}(a), (b) show RHEED images acquired along the [100] direction of Sr$_3$SnO and Sr$_3$PbO, respectively. We note that the underlying YSZ substrate and a thin SrO buffer layer deposited prior to the antiperovskite are also cubic with similar lattice constants. However, they are forbidden by their crystal structure from exhibiting (0$l$) streaks with odd integer $l$. Hence, the appearance of the (01) streak establishes the existence of the target antiperovskite phase.

Figures~\ref{Fig2}(c)-(e) and (f)-(h) show LEED images acquired at different energies for Sr$_3$SnO and Sr$_3$PbO, respectively. The square array of the diffraction spots is consistent with the antiperovskite crystal structure. In addition, the complex evolution of the structure factor as a function of electron energy is observed~\cite{vanHove_1986}. In general, LEED images of Sr$_3$PbO exhibit brighter patterns than those of Sr$_3$SnO, and this is also reflected in the RHEED streaks. As discussed in the following two subsections, XPS measurements show that both the Sr$_3$SnO and Sr$_3$PbO films have a thin surface layer covering the bulk antiperovskite phase. The brighter RHEED/LEED images in Sr$_3$PbO may point to a thinner surface layer covering Sr$_3$PbO, or to the stronger scattering strength of Pb compared to Sn.

\subsection{Sr$_3$SnO XPS}

\begin{figure}
\includegraphics[scale=1]{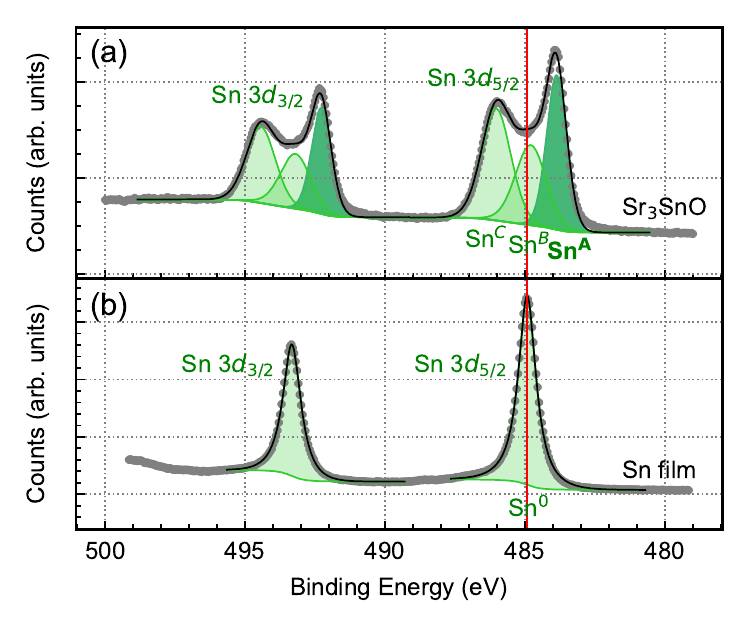}
\caption{(a) XPS spectrum of the Sn 3$d_{3/2}$ and $3d_{5/2}$ doublet of Sr$_3$SnO (sample SS91), acquired at MPI-FKF with photon energy 1486.6 eV. The gray circles are the measured data, the black line is the overall fit and the green shaded areas are the individual peaks that constitute the fit. (b) A reference spectrum for a thin film of metallic Sn is shown for comparison with (a).}
\label{Fig3}
\end{figure}

Figure~\ref{Fig3}(a) presents the Sn $3d$ spectrum of a Sr$_3$SnO film (SS91). As clearly seen, each of the spin-split levels ($3d_{3/2}$ and $3d_{5/2}$) exhibits two pronounced peaks, indicative of multiple Sn valence states. Using the fitting procedure described in Section~\ref{SecMet}, we find that actually, a minimum of three Gaussian-Lorentzian mixture functions are required to fit each level. For the Sn 3$d_{5/2}$ level, the three peaks are centered at 483.87 eV, 484.79 eV and 486.03 eV [Table~\ref{T_SS}]. We label these peaks as Sn$^A$, Sn$^B$ and Sn$^C$, respectively [Fig.~\ref{Fig3}(a)].

\begin{table}[h]
\center
\setlength{\tabcolsep}{7pt}
\caption{Binding energies of peaks extracted from fits to the Sn $3d_{5/2}$ level of Sr$_3$SnO. As described in the main text, Sn$^A$ is assigned to the bulk antiperovskite phase, whereas Sn$^B$ and Sn$^C$ are ascribed to a surface layer present in the films. The binding energy of Sn$^0$, extracted from a reference Sn film, is also reported.}
\begin{tabular}{c|ccc}
\hline 
 & Sn$^A$ [eV] & Sn$^B$ [eV] & Sn$^C$ [eV] \\
\hline\hline
SS60 [Fig.~\ref{Fig4}(b)] & 483.82 & 484.72 & 486.02 \\
SS91 [Fig.~\ref{Fig3}(a)] & 483.87 & 484.79 & 486.03 \\
\hline
\multicolumn{4}{c}{} \\
\hline 
 & Sn$^0$ [eV] & & \\
\hline\hline
Sn film [Fig.~\ref{Fig3}(b)] & 484.92 \\
\hline
\end{tabular}
\label{T_SS}
\end{table}

To understand the origin of these peaks, we performed XPS measurements on a control sample, a thin film of Sn deposited on YSZ [Fig.~\ref{Fig3}(b)]. The Sn 3$d_{5/2}$ level shows a sharp peak centered at 484.92 eV, closely matching the literature value for metallic Sn, 485.0 eV~\cite{Moulder_1992}; we thus label this peak Sn$^0$. Comparing with the spectrum from Sr$_3$SnO [Fig.~\ref{Fig3}(a)], we note that Sn$^0$ overlaps with Sn$^B$. Sn$^C$, with higher binding energy, matches literature values for SnO~\cite{Moulder_1992}, or SnO$_2$~\cite{Themlin_PRB_1992}. Sn$^A$, with lower binding energy, could be assigned to the antiperovskite phase. Intuititively, Sn states with higher binding energy than Sn$^0$ (red line in Fig.~\ref{Fig3}) are cationic (positively charged), such that core electrons are less readily removed, whereas Sn states with lower binding energy than Sn$^0$ are anionic (negatively charged), such that core electrons are more readily removed. Indeed, XPS measurements of Ni$_3$Sn$_4$ electrodes for Li-ion batteries showed that when Sn was lithiated and therefore negatively charged, the 3$d_{5/2}$ peak corresponding to Sn$^0$ shifted to lower binding energies~\cite{Ehinon_ChemMater_2008}.

\begin{figure}
\includegraphics[scale=1]{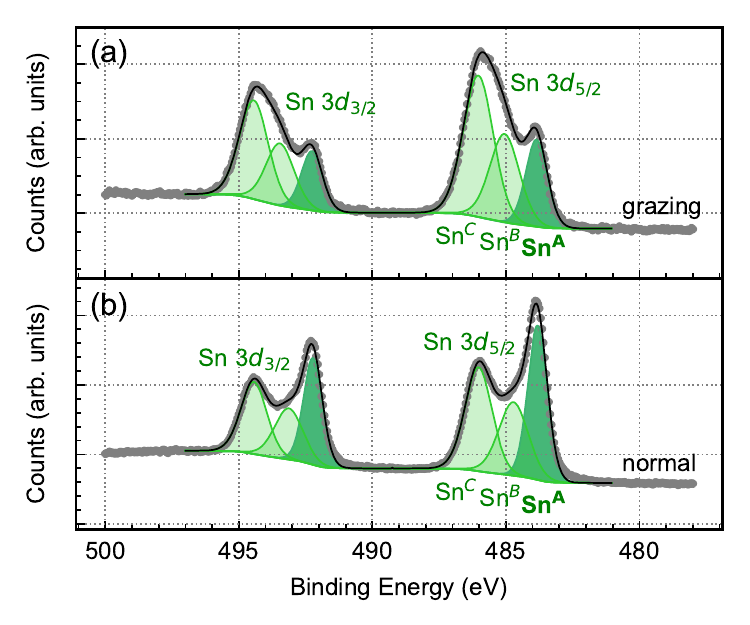}
\caption{Angle dependence: Sn $3d$ spectra of sample SS60, taken at (a) grazing (60$^{\circ}$ off normal) and (b) normal emission. Data were acquired at MPI-FKF with photon energy $h\nu$ = 1486.6 eV. }
\label{Fig4}
\end{figure}

The dependence of the Sn $3d$ spectra on the emission angle of the electrons is shown in Fig.~\ref{Fig4} for another Sr$_3$SnO film (SS60). The spectrum obtained at grazing emission is more surface sensitive than that at normal emission. We observe that near the surface, Sn$^B$ and Sn$^C$ occupy a greater fraction of the total intensity than Sn$^A$ [Fig.~\ref{Fig4}(a)]. Deeper into the bulk, however, Sn$^A$ is enhanced relative to Sn$^B$ and Sn$^C$ [Fig.~\ref{Fig4}(b)]. Thus, the bulk phase of our film is characterized by Sn$^A$, consistent with anionic Sn in Sr$_3$SnO. Nevertheless, there is a surface layer in which Sn reverts to its neutral (Sn$^B$ $\sim$ Sn$^0$) and cationic (Sn$^C$ $\sim$ SnO or SnO$_2$) states, likely originating from the unstable polar (001) surface of Sr$_3$SnO. Since an electron with kinetic energy on the order of 1 keV has an inelastic mean free path on the order of 1 nm~\cite{Powell_JVSTA_1999}, we deduce the surface layer to have thickness less than 1 nm.

\subsection{Sr$_3$PbO XPS}

\begin{figure}
\includegraphics[scale=1]{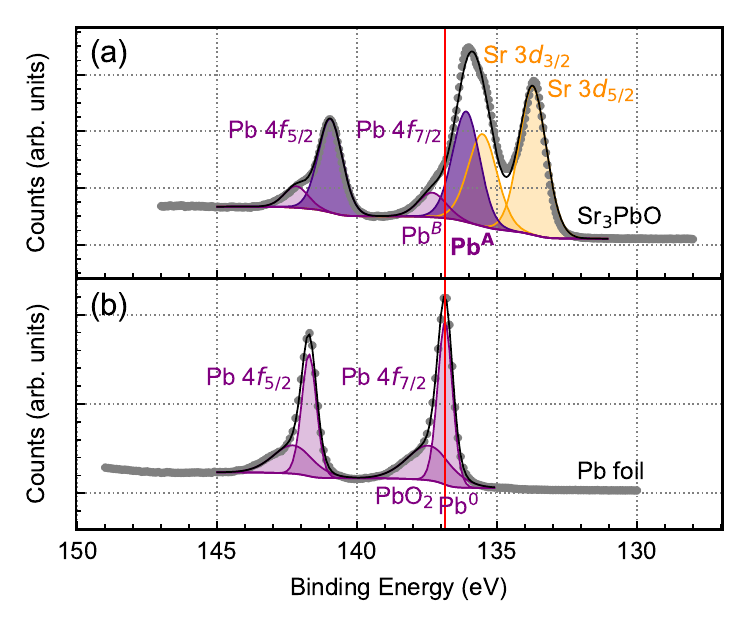}
\caption{(a) XPS spectrum of the Sr 3$d_{3/2}$ and 3$d_{5/2}$ doublet and the Pb 4$f_{5/2}$ and 4$f_{7/2}$ doublet of Sr$_3$PbO (sample AP337), acquired at MPI-FKF with photon energy 1486.6 eV. The gray circles are the measured data, the black line is the overall fit and the orange (purple) shaded areas are the individual Sr (Pb) peaks that constitute the fit. (b) A reference spectrum for a metallic Pb foil is shown for comparison with (a).}
\label{Fig5}
\end{figure}

In essence, the XPS results of Sr$_3$PbO are similar to the Sr$_3$SnO results. Figure~\ref{Fig5}(a) presents the overlapping Sr $3d$ and Pb $4f$ spectra of a Sr$_3$PbO film (AP337). Two Gaussian-Lorentzian mixture functions were required to fit the Pb 4$f$ levels, with peaks Pb$^A$ = 136.10 eV and Pb$^B$ = 137.32 eV [Table~\ref{T_AP}]. To identify these peaks, we again performed XPS measurements on a control sample, a Pb foil cleaned \textit{in situ} by Ar sputtering [Fig.~\ref{Fig5}(b)]. The Pb 4$f_{7/2}$ level shows a pronounced peak centered at 136.86 eV, close to the literature value for metallic Pb, 136.9 eV~\cite{Moulder_1992}; we thus label this peak Pb$^0$. There is also a residual peak at higher binding energies, 137.43 eV, which agrees with literature values for PbO$_2$~\cite{Moulder_1992}. Comparing with the data from Sr$_3$PbO [Fig.~\ref{Fig5}(a)], we observe that Pb$^B$ overlaps with PbO$_2$, whereas Pb$^A$ is exclusive to the antiperovskite film. Its lower binding energy relative to Pb$^0$ indicates that it is anionic.

\begin{table}[h]
\center
\setlength{\tabcolsep}{7pt}
\caption{Binding energies of peaks extracted from fits to the Pb $4f_{7/2}$ level of Sr$_3$PbO. As described in the main text, Pb$^A$ is assigned to the bulk antiperovskite phase, whereas Pb$^B$ is ascribed to a surface layer present in the films. The binding energies of Pb$^0$ and PbO$_2$, extracted from a reference Pb foil, are also reported.}
\begin{tabular}{c|cc}
\hline 
 & Pb$^A$ [eV] & Pb$^B$ [eV] \\
\hline\hline
AP149 [Fig.~\ref{Fig6}(b)] & 136.10 & 137.42  \\
AP337 [Fig.~\ref{Fig5}(a)] & 136.10 & 137.32 \\
\hline
\multicolumn{3}{c}{} \\
\hline 
 & Pb$^0$ [eV] & PbO$_2$ [eV] \\
\hline\hline
Pb foil [Fig.~\ref{Fig5}(b)] & 136.86 & 137.43 \\
\hline
\end{tabular}
\label{T_AP}
\end{table}

\begin{figure}
\includegraphics[scale=1]{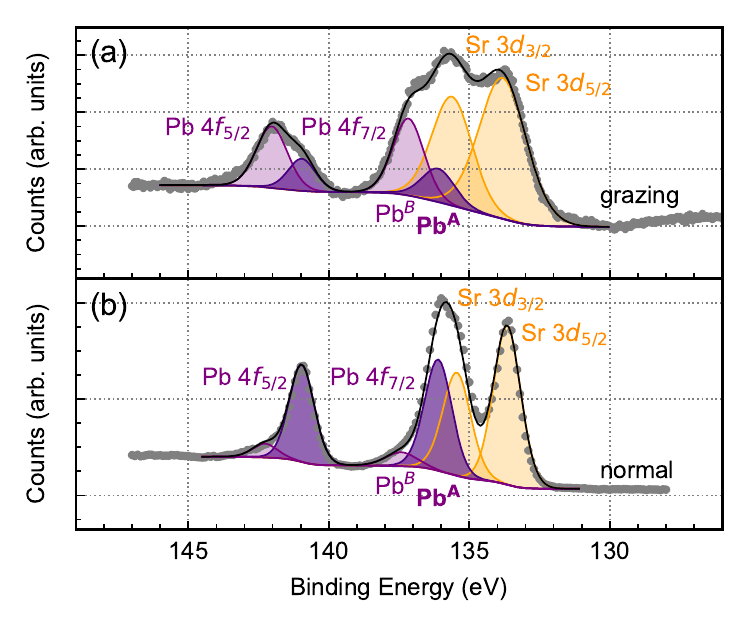}
\caption{Angle dependence: Sr 3$d$ and Pb 4$f$ spectra of sample AP149, taken at (a) grazing (60$^{\circ}$ off normal) and (b) normal emission. Data were acquired at MPI-FKF with photon energy $h\nu$ = 1486.6 eV.}
\label{Fig6}
\end{figure}

Figure~\ref{Fig6} presents the angle dependence of the overlapping Sr 3$d$ and Pb 4$f$ spectra, along with fits, for sample AP149. At grazing emission, Pb$^B$ dominates the spectrum, but at normal emission, the intensity of Pb$^A$ is enhanced relative to Pb$^B$. Again, we conclude that the bulk phase of our film is characterized by Pb$^A$, which we assign to anionic Pb in Sr$_3$PbO, but in a thin surface layer, Pb reverts to its cationic state (Pb$^B$ $\sim$ PbO$_2$).

At the given photon energy of 1486.6 eV, with normal emission, the XPS signal from the surface layer appears to be less significant in Sr$_3$PbO [Figs.~\ref{Fig5}(a),~\ref{Fig6}(b)] than in Sr$_3$SnO [Figs.~\ref{Fig3}(a),~\ref{Fig4}(b)]. This could be attributed to the difference in kinetic energy of the emitted electron: the higher binding energy for the Sn 3$d$ core levels relative to the Pb 4$f$ core levels may cause the XPS signal for Sr$_3$PbO to be more bulk sensitive. As a result, it is possible that neutral Pb exists in the surface layer of Sr$_3$PbO, similar to Sr$_3$SnO, but the signal is too small to detect. Alternatively, the surface layer in Sr$_3$PbO could simply be thinner. It is possible that anionic Pb is intrinsically more stable and less prone to oxidation than anionic Sn. 

\section{Density functional theory}
\label{secDFT}

In this section, we use DFT to demonstrate that in the antiperovskite compounds $A_3B$O, $B$ (= Sn, Pb) does indeed carry a negative effective charge, consistent with the heuristic concept of formal oxidation states. Then we show that the XPS binding energy of the Sn$^A$ (Pb$^A$) peak in Sr$_3$SnO (Sr$_3$PbO) relative to the Sn$^0$ (Pb$^0$) peak in metallic Sn (Pb) matches predictions by DFT calculcations. This further confirms our assignment of the Sn$^A$ and Pb$^A$ peaks to the bulk antiperovskite phase at a quantitative level.

\subsection{Effective charges}

While formal oxidation states are a useful construct when examining chemical bonding or electronic structure, they are not identical to the actual effective charges surrounding each atom. Generally, bonds in a crystal exhibit a greater degree of covalency than is expected in a pure ionic picture. To compute the effective charges, we used Bader's method of partitioning the charge density via zero-flux surfaces~\cite{Tang_JPCM_2009}. Table~\ref{T_Bader} presents effective charges computed for various antiperovskites. The effective charge of $B$ (= Sn, Pb) averages around $-2$ across the compounds considered. Thus, while the effective charge is clearly lower than the value of $-4$ expected from formal oxidation states, it is still clear that $B$ is unusally anionic. We note a trend that as the size of $A$ increases from Ca to Sr to Ba, the effective charge of $B$ becomes less negative~\cite{Kariyado_PRM_2017}.

\begin{table}[h]
\center
\setlength{\tabcolsep}{7pt}
\caption{Effective charges computed by Bader analysis for various antiperovskites $A_3B$O.}
\begin{tabular}{c|ccc}
\hline 
 & $A$ = Ca/Sr/Ba & $B$ = Sn/Pb & O \\
\hline\hline
Ca$_3$SnO & $+1.30$ & $-2.38$ & $-1.51$ \\
Ca$_3$PbO & $+1.29$ & $-2.35$ & $-1.52$ \\
Sr$_3$SnO & $+1.26$ & $-2.30$ & $-1.48$ \\
Sr$_3$PbO & $+1.25$ & $-2.26$ & $-1.48$ \\
Ba$_3$SnO & $+1.14$ & $-1.98$ & $-1.44$ \\
Ba$_3$PbO & $+1.12$ & $-1.93$ & $-1.44$ \\
\hline
\end{tabular}
\label{T_Bader}
\end{table}

\subsection{Core-level shifts}

Experimentally, what XPS measures is neither the formal oxidation state nor the effective charge, but shifts in the core-level binding energies. We therefore used DFT to quantitatively confirm the shift towards lower binding energies for anionic $B$ relative to metallic $B$. To calculate the core-level binding energies ($E_c$), we worked within the initial state approximation, wherein a selected core electron is removed, but the remaining electrons are kept frozen~\cite{Kohler_PRB_2004}. Then $E_c$ is simply given by the Kohn-Sham (KS) eigenvalue of the core electron ($\epsilon_c$), relative to the Fermi energy ($\epsilon_F$):
\begin{equation}
E_c = -(\epsilon_c - \epsilon_F).
\end{equation}
While final state effects, primarily the screening of the core hole, are neglected, the initial state approximation captures the chemical state of the atom as reflected in its valence charge configuration~\cite{Bellafont_PCCP_2015}. 

In the PAW formalism of \texttt{VASP}, $\epsilon_c$ is computed in two steps~\cite{Kohler_PRB_2004}: First, the core electrons are frozen and the valence charge density is computed via the normal, self-consistent electronic relaxation. Second, the KS eigenvalues for the core electrons are solved inside the PAW spheres while keeping the valence charge density fixed. As a check, we also performed all-electron PBE GGA calculations using
the relativistic linear muffin-tin orbital (LMTO) method as implemented in the \texttt{PY LMTO} computer code. Here, spin-orbit coupling was included by solving the Dirac equations inside the atomic spheres. Some details of the implementation can be found in Ref.~\cite{Antonov_2004}.

The absolute values of $E_c$ as determined from the KS eigenvalues of DFT are typically 20-30 eV lower than the experimental values reported by XPS, due to a breakdown of Koopman's theorem~\cite{vanSetten_JCTC_2018}. However, DFT does provide meaningful values of $\Delta E_c$, the shift of the core-level binding energy between two systems. In a study, van Setten \textit{et al.} demonstrated that for a set of molecules containing C, O, N or F, the mean absolute difference between $\Delta E_c$, as calculated from the KS eigenvalues, and the actual core-level shifts, as measured by XPS, was only 0.74 eV~\cite{vanSetten_JCTC_2018}.

To make a meaningful comparison with our data, we took metallic Sn ($\alpha$ allotrope, diamond structure) and metallic Pb (face-centered cubic) as our references. We calculated $\Delta E_{\textrm{Sn }3d}$ between Sr$_3$SnO and $\alpha$-Sn, and $\Delta E_{\textrm{Pb }4f}$ between Sr$_3$PbO and Pb. Dense BZ sampling and a high energy cutoff, as stated previously, were needed to converge the core-level binding energies within 1 meV. The results are shown in Table~\ref{T_CLS}, computed using the following experimental lattice constants: 5.139 \AA~for Sr$_3$SnO, 6.489 \AA~for $\alpha$-Sn, 5.151 \AA~for Sr$_3$PbO and 4.950 \AA~for Pb~\cite{Nuss_ACSB_2015, Thewlis_Nature_1954, Bouad_JSSC_2003}. We note that in the case of $\Delta E_{\textrm{Sn }3d}$, there is a shift by $+0.12$ eV when the DFT-optimized lattice parameter is used, due to a discrepancy of 2.5\% between the experimental and DFT-optimized lattice constants of $\alpha$-Sn. We also note that differences in $\Delta E_c$ arising from the use of the local density approximation (LDA) instead of GGA are within 0.1 eV.

\begin{table}
\center
\setlength{\tabcolsep}{12pt}
\caption{Comparison between DFT and XPS. The core-level shifts, as predicted by DFT, are given by $\Delta E_{\textrm{Sn }3d} = E_{\textrm{Sn }3d}(\textrm{Sr}_3\textrm{SnO}) - E_{\textrm{Sn }3d}(\alpha\textrm{-Sn})$ and $\Delta E_{\textrm{Pb }4f} = E_{\textrm{Pb }4f}(\textrm{Sr}_3\textrm{PbO}) - E_{\textrm{Pb }4f}(\textrm{Pb})$. Results from two different codes (\texttt{VASP} and \texttt{PY LMTO}) are shown. The core-level shifts, as measured by XPS, are given by the difference between the Sn$^A$ peak in Sr$_3$SnO and the Sn$^0$ peak in metallic Sn, or between the Pb$^A$ peak in Sr$_3$PbO and the Pb$^0$ peak in metallic Pb.}
\begin{tabular}{c|c}
\hline 
DFT: $\Delta E_{\textrm{Sn }3d}$ [eV] & DFT: $\Delta E_{\textrm{Pb }4f}$ [eV] \\
\hline
\texttt{VASP}: $-0.95$ & \texttt{VASP}: $-0.79$ \\
\texttt{PY LMTO}: $-1.14$ & \texttt{PY LMTO}: $-0.98$ \\
\hline\hline
XPS: Sn$^A$ $-$ Sn$^0$ [eV] & XPS: Pb$^A$ $-$ Pb$^0$ [eV] \\
\hline
SS60: $-1.10$ & AP149: $-0.76$ \\
SS91: $-1.05$ & AP337: $-0.76$ \\
\hline
\end{tabular}
\label{T_CLS}
\end{table}

Shown in Table~\ref{T_CLS} are also the core-level shifts as measured by XPS. For the Sr$_3$SnO films, we took the difference between the Sn$^A$ peak, which we ascribed to the bulk antiperovskite phase, and the Sn$^0$ peak in the reference Sn metal. Similarly, the difference between Pb$^A$ in Sr$_3$PbO and Pb$^0$ in Pb metal was used to derive the shift for Sr$_3$PbO. The DFT and XPS results for the core-level shifts show a very good agreement. Furthermore, in both theory and experiment, the magnitude of the shift is larger in Sn 3$d$ compared to Pb 4$f$. Hence, we conclude again with additional confirmation that the Sn$^A$ (Pb$^A$) peak corresponds to anionic Sn (Pb) in Sr$_3$SnO (Sr$_3$PbO).

\section{Summary}

In this work, we have investigated the antiperovskites Sr$_3$SnO and Sr$_3$PbO, whose predicted topological crystalline insulating phase and approximate 3D Dirac semimetallic phase hinge upon the stabilization of Sn and Pb in an unusual anionic state ($\sim$$-2$ according to Bader charge analysis). Our XPS measurements, along with DFT calculations, confirm that anionic Sn and Pb do indeed exist in thin films of Sr$_3$SnO and Sr$_3$PbO. Interestingly though, we observed signatures of cationic or neutral Sn and Pb distributed at the surface of the films. This suggests that the polar (001) surface of these antiperovskites is susceptible to a reconstruction wherein the valence states of Sn and Pb are altered. Such a modification is likely to have drastic impact on the surface electronic structure. We suggest using scanning tunneling microscopy to elucidate the nature of potential surface reconstruction (electronic or atomic) and its effects on putatitve topological surface states. 

\begin{acknowledgments}

We thank U. Wedig for helpful discussions. We also thank M. Konuma, C. M\"{u}hle, K. Pflaum, S. Prill-Diemer and S. Schmid for technical assistance at MPI-FKF. We acknowledge the Paul Scherrer Institut, Villigen, Switzerland for provision of synchrotron radiation beamtime at the ADRESS beamline of the SLS. D. H. acknowledges support from a Humboldt Research Fellowship for Postdoctoral Researchers. N. B. M. S acknowledges partial financial support from Microsoft.

\end{acknowledgments}


%

\end{document}